# Stable compounds in the $CaO-Al_2O_3$ system at high pressures

Ekaterina I. Marchenko[1,2], Artem R. Oganov[3], Efim A. Mazhnik[3], Nikolay N. Eremin[1]

[1] – Department of Geology, Lomonosov Moscow State University, GSP-1, 1 Leninskiye Gory, Moscow 119991, Russia.
[2] – Laboratory of New Materials for Solar Energetics, Faculty of Materials Science, Lomonosov Moscow State University, Moscow, Russia
[3] – Skolkovo Institute of Science and Technology, 30 bldg. 1, Bolshoy blvd., Moscow 121205, Russia.

**Keywords:** $CaO-Al_2O_3$ system, crystal structure prediction, high pressure, evolutionary algorithm, USPEX, Earth's mantle

## Abstract

Using evolutionary crystal structure prediction algorithm USPEX, we showed that at pressures of the Earth's lower mantle $CaAl_2O_4$ is the only stable calcium aluminate. At pressures above 7.0 GPa it has the $CaFe_2O_4$-type structure and space group *Pnma*. This phase is one of prime candidate aluminous phases in the lower mantle of the Earth. We show that at low pressures $5CaO \bullet 3Al_2O_3$ ($C_5A_3$) with space group $Cmc2_1$, $CaAl_4O_7$ ($C2/c$) and $CaAl_2O_4$ ($P2_1/m$) structures are stable at pressures of up to 2.1, 1.8 and 7.0 GPa respectively. The previously unknown structure of the orthorhombic "CA-III" phase is also found from our calculations. This phase is metastable and has a layered structure with space group $P2_12_12$.

## Introduction

Among the most common chemical elements of the Earth's crust and mantle, aluminum ranks fifth Al/Si ratio being 0.36 in the crust and 0.17 in the mantle [1]. Today, it is thought that the lower mantle consists mainly of (Mg,Fe)$SiO_3$ bridgmanite with a perovskite-type (Pv) crystal structure (~70 vol.%), (Mg,Fe)O ferropericlase (~20 vol.%), and calcium perovskite $CaSiO_3$ (Ca-Pv) (6-12 vol.%) [2-4]. It is not entirely clear whether Al forms its own minerals or is present as an impurity in silicates in the lower mantle [5]. According to the predominant view, the main reservoir of Al in the mantle is not its own phases, but Pv with several percent of Al impurity. Three possible mechanisms of Al atoms entering the Pv [6,7] and $MgSiO_3$ post-perovskite structures [8] need to be considered:

1) Mg+Si → Al+Al (CCM – coupled charge mechanism), maintaining local charge balance,
2) Si+Si+O → Al+Al+$V_O$ (OVM – oxygen vacancy mechanism), with a vacancy in the oxygen site to maintain the charge balance,
3) Si → Al+H (AlHM – aluminum hydrogen mechanism), with a $H^+$ to maintain charge balance.

Previous studies [9,10] have shown that Pv can contain up to 10% $Al_2O_3$ via the CCM, whereas the isomorphic capacity of Ca-Pv for aluminum is low. Theoretical calculations [6,7,11] have confirmed the preference of the CCM mechanism for the entry of Al in Pv and shown that the OVM is the preferred mechanism for the incorporation of aluminum in Ca-Pv. According to [11], only part of the total amount of Al in the mantle can be concentrated in Pv and Ca-Pv phases and consequently Al might form its own phases in the mantle, the simplest options being $CaAl_2O_4$ and $MgAl_2O_4$. Earlier it was shown that another candidate stoichiometry, $Al_2SiO_5$, is unstable at pressures of the lower mantle [12].

All experimentally known polymorphic modifications of $CaAl_2O_4$ can be divided into low-pressure structures, in which aluminum atoms are in the tetrahedral coordination, and high-pressure phases, with aluminum in 5- or 6-fold coordination. The well-known monoclinic modification with the space group $P2_1/n$ at normal conditions [13] consist of a framework of $AlO_4$ tetrahedra, topologically equivalent to the tridymite structure ($SiO_2$) (Figure 1a). The voids of the framework are occupied by calcium atoms in three independent crystallographic sites with different coordination numbers: 6 (Ca1), 6 (Ca2) and 9 (Ca3). This modification was first observed in meteorites and the mineral was named crotite [14]. Another monoclinic modification of $CaAl_2O_4$ with the space group $P2_1/c$ (Figure 1b) crystallizes in the *m*-$CaGa_2O_4$ structure type [13], where the $AlO_4$ distorted tetrahedra form a three-dimensional framework with the tridymite topology, and Ca atoms are seven-coordinate. This phase was also found in meteorites and was named dmitriivanovite [15]. The monoclinic $P2_1/c$ phase is slightly denser than $P2_1/n$, and the transition from $P2_1/n$ to $P2_1/c$ is observed at the pressure of 1 GPa and the temperature of 700 °C and at 2 GPa and 1300 °C [13]. Among other low-pressure modifications of $CaAl_2O_4$ with $AlO_4$ tetrahedra, a crystal structure with space group $P6_3$ has been mentioned [16]. However, the authors have questioned the correctness of their structure refinement because of the high R-factor and highly distorted interatomic distances. Most likely, this phase is metastable at low pressures and does not have a stability field in the phase diagram.

With further increase in pressure, the coordination of Al atoms changes from tetrahedral to octahedral. Among the $CaAl_2O_4$ phases with the octahedral coordination of Al atoms the best known orthorhombic *Pnma* modification was first described in 1957 [17] (Figure 1c). This phase crystallizes in the $CaFe_2O_4$ structure type (space group *Pnma*) where edge-connected $AlO_6$ octahedra form channels along the c axis, filled with $Ca^{2+}$ cations.

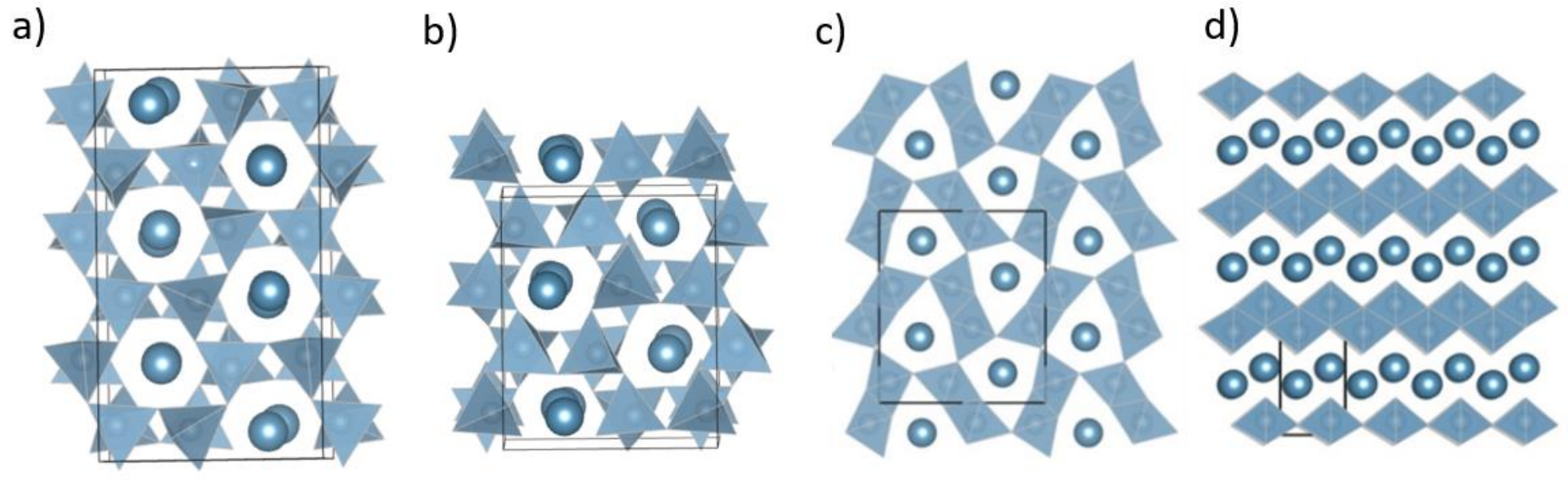

Figure 1. Polyhedral representation of experimentally known $CaAl_2O_4$ structures with $AlO_4$ tetrahedra: $P2_1/n$ (*xz* projection), $P2_1/c$ (*yz* projection); and $AlO_6$ octahedra: a) *Pnma* modification (*xy* projection) and b) "layered" $P2_1/m$ modification (*zy* projection).

Another less dence modification of $CaAl_2O_4$ is the monoclinic $P2_1/m$ phase with a layered structure, synthesized by Lazić et al. [18](Figure 1d). The layers of $AlO_6$ octahedra in this structure are separated by interlayer Ca cations. The phase transition $P2_1/m \rightarrow Pnma$ has been experimentally discovered at $P$ = 4-8 GPa [18]. In the experimental study of the $CaAl_2O_4$-$CaGa_2O_4$ system [13], yet another modification of $CaAl_2O_4$, named "CA-III", and stable in the pressure range of 3-8 GPa, was described. Its crystal structure was not determined, but an orthorhombic cell was established with the parameters $a$=4.39 Å, $b$=5.07 Å, $c$=6.96 Å.

In a recent theoretical paper [19], it has been shown that the phase transition between $P2_1/m$ and *Pnma* is predicted by ab initio calculations at 7-8 GPa, which is in close agreement with experiment [18]. Calculations based on semiclassical interatomic potentials overestimate of the phase transition pressure, 18-19 GPa [19]. As calculations [19] showed, when pressure increases, *Pnma*-$CaAl_2O_4$ increasingly more preferred in the entire range of mantle pressures and temperatures. The *Cmcm* phase of $CaAl_2O_4$ is always slightly (0.2-0.5 eV/f.u.) less stable than the *Pnma* phase.

Analyzing recent experimental and theoretical studies, one can conclude that under *P-T* conditions of the Earth's mantle, only two phases of $CaAl_2O_4$ may exist: monoclinic $P2_1/m$ and orthorhombic *Pnma*, both with the octahedral coordination of the Al atoms, and the stability field of the latter covers almost the entire P-T range of the Earth's mantle. So far, only the phases of the stoichiometry $CaAl_2O_4$ have been investigated at conditions of the Earth's mantle, and even there some structures (such as Ca-III) remain experimentally unresolved.

In addition to the structural modifications of $CaAl_2O_4$, the literature describes experimentally investigated phases with different stoichiometric ratios where the aluminum atoms have tetrahedral coordination ($Ca_{12}Al_{14}O_{33}$ [20], $CaAl_4O_7$ [21], $Ca_3Al_2O_6$ [22], $Ca_5Al_6O_{14}$ [23], $Ca_4Al_6O_{13}$ [24]. The experimentally known phases in which aluminum atoms in mixed coordination have been described: $Ca_2Al_2O_5$ [25,26] with aluminum in tetrahedra and octahedra; $Ca_4Al_6O_{13}$ [27], with aluminum in tetrahedra, octahedra and trigonal bipyramids; $Ca_3Al_2O_6$ [28] with Al in octahedra and squares, and $CaAl_{12}O_{19}$ [29] with aluminum in octahedra and trigonal bipyramids. The mineral mayenite [30], which has a clathrate structure, has a rather complex composition $Ca_{12}Al_{14}O_{33}$ and removal of one oxygen atom per formula unit leads to a peculiar metallic electride state [31] (Figure S1). Overall, the tetrahedral coordination of aluminum is unstable at pressures above 2-8 GPa [5,13,19], therefore, phases with tetrahedral coordination of Al atoms in the CaO-$Al_2O_3$ system cannot exist in the deep mantle.

In this work we theoretically predicted thermodynamically stable phases in the CaO-$Al_2O_3$ system at pressures in the range 0-200 GPa using the Universal Structure Predictor: Evolutionary Xtallography (USPEX) algorithm [32,33].

## Methods

Evolutionary searches for stable phases in the CaO-$Al_2O_3$ system were performed using the USPEX code [32-34] in its variable-composition mode of USPEX. In our evolutionary searches, all structures and stoichiometries were allowed with up to 32 atoms in the primitive cell. The first generation contained 200 structures and was produced using symmetric [*Lyakhov et al.,* 2013] and topological random structure generators [35]. All subsequent generations had 120 structures. To determine stable compounds, USPEX builds convex hull diagrams (also known as Maxwell construction). Points lying on the hull indicate thermodynamically stable phases. Points lying above the hull are thermodynamically unstable phases. Evolutionary searches were performed at pressures of 25, 50, 100, 200 GPa. All structures were relaxed using VASP [36], using density functional theory at the level of generalized gradient approximation (PBE exchange-correlation functional [37] and PAW method [38,39], with plane-wave cutoff of 600 eV. Brillouin zone was sampled using uniform k-meshes with reciprocal-space resolution $2\pi \times 0.05$ Å$^{-1}$. Low-enthalpy structures predicted using USPEX and experimentally known structures of different stoichiometries were then carefully relaxed at many pressures in the range 0-200 GPa, and a full zero-temperature pressure-composition phase diagram of the CaO-$Al_2O_3$ system was built. Crystal structures were visualized using the VESTA program [40].

## Results and discussion

Variable-composition USPEX searches allow for a detailed sweep of the entire compositional space in a single simulation. They recovered known stable structures and phase transitions between them in CaO [41,42] and $Al_2O_3$ [43,44].

Interestingly, USPEX has also found (as stable or metastable at zero Kelvin) all previously known structures of $CaAl_2O_4$ with the octahedral coordination of aluminum atoms, including the marokite-type phase and the "layered" $P2_1/m$ modification. As we find, the *Pnma* modification of $CaAl_2O_4$ with a marokite structure type turns out to be stable at pressures from 7.0 to at least 200 GPa (Figures 2 and 3). At pressures below 7.0 GPa, the $P2_1/m$ structure is stable, and below 1.1 GPa the $P2_1/n$ structure is stable (Figure 2). The $P2_1/c$ phase is unstable over the entire pressure range. At low pressures, we see yet another two phases as stable: 5CaO • $3Al_2O_3$ ($C_5A_3$) with space group $Cmc2_1$ (up to 2.1 GPa) and $CaAl_4O_7$ with space group $C2/c$ (up to 1.8 GPa) (Figure 3a). However, the stability of $C_5A_3$ phase is marginal: if we include zero-point energy (we did its crude estimate with Debye model), $C_5A_3$ phase is no longer thermodynamically stable, but is 0.009 eV/block above the convex hull. The convex hull of the CaO-$Al_2O_3$ system at pressures of 0 GPa and 100 GPa are shown in Figure 4. It should be noted that the $Ca_3Al_2O_6$ ($C_3A$) and $CaAl_{12}O_{19}$ ($CA_6$) and $Ca_{12}Al_{14}O_{33}$ ($C_{12}A_7$) phases at low pressures are very close to the convex hull, being just 0.008, 0.027, and 0.008 eV/block, respectively, above it at pressure of 0 GPa. The calculated fitness (the height above the convex hull, characterizing the degree of instability of a given structure) for these experimental structures as a function of pressure is shown of Figure S3.

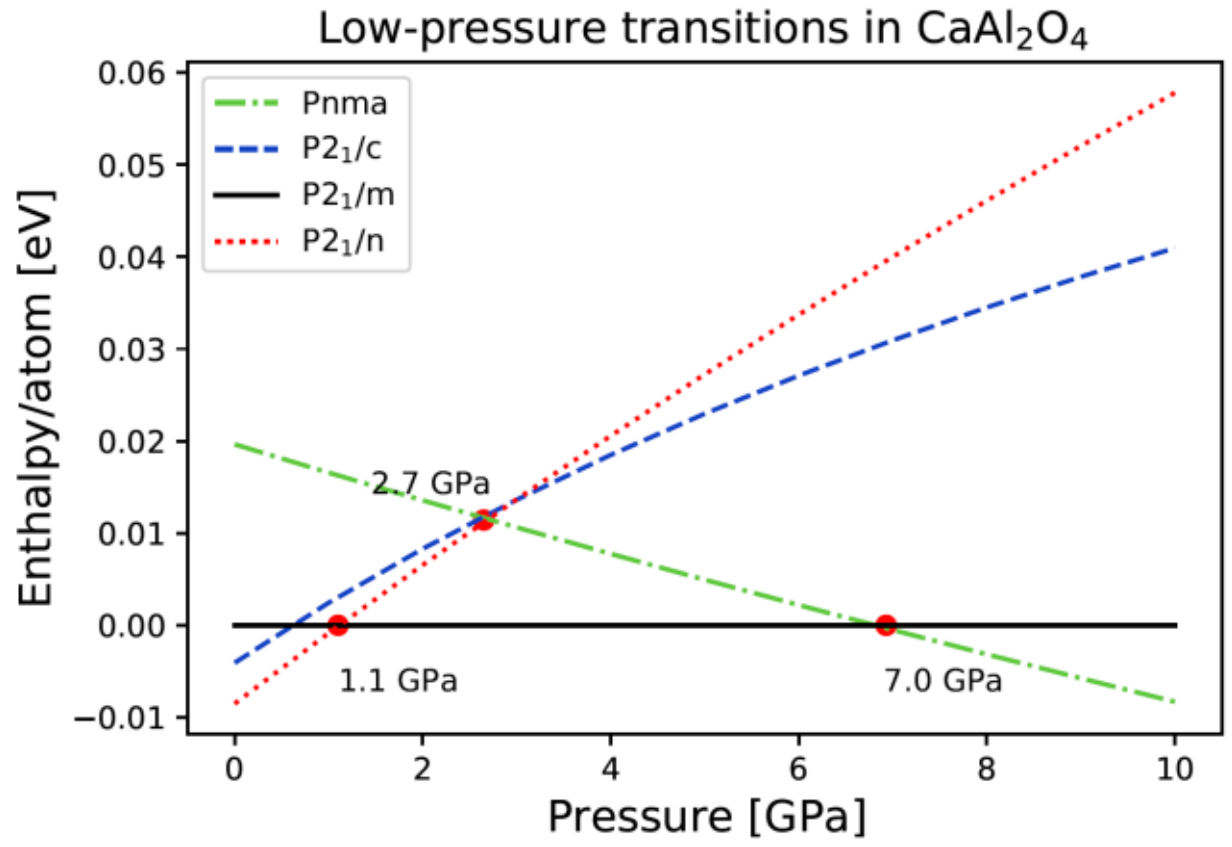

Figure 2. Enthalpy differences between low-pressure phases of $CaAl_2O_4$. Enthalpy for $P2_1/m$ phase was taken as a reference.

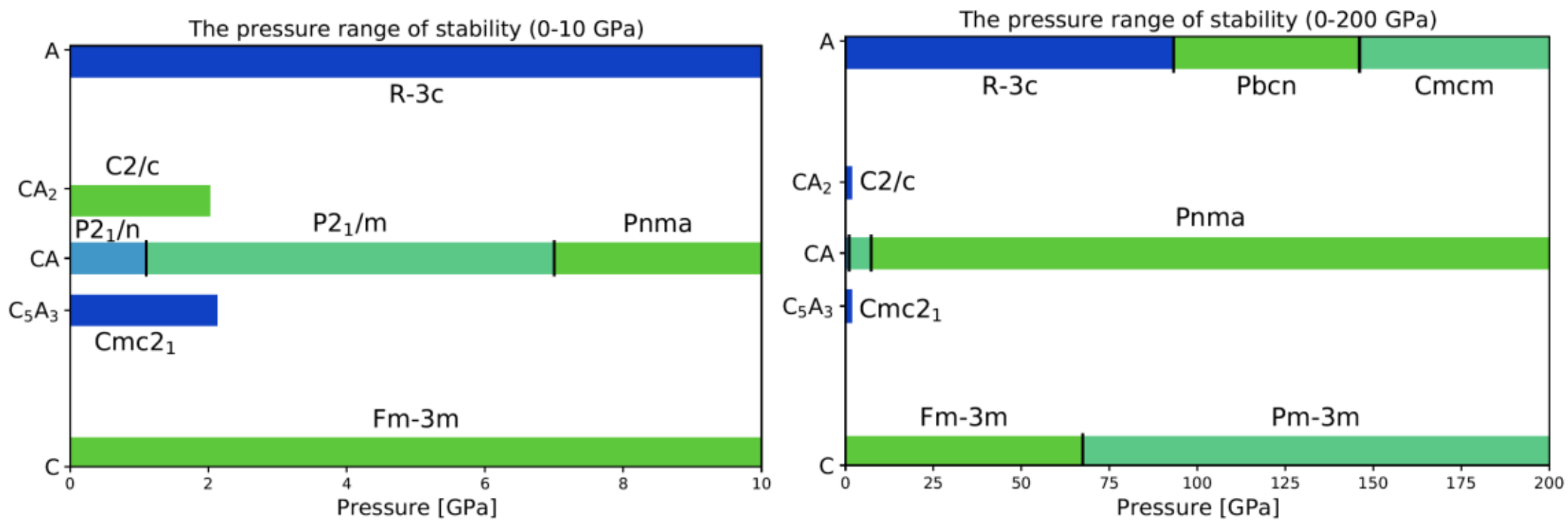

a) b)

Figure 3. Pressure-composition phase diagram of the $CaO$-$Al_2O_3$ system (a) in the 0-10 GPa pressure range, (b) in the 0-200 GPa pressure range.

We also calculated phase polymorphic transitions for CaO and $Al_2O_3$. For CaO, the pressure of the phase transition from $Fm\bar{3}m$ to $Pm\bar{3}m$ is 67.4 GPa (Figure S2a). For $Al_2O_3$, the pressure of phase transitions from $R\bar{3}c$ to *Pbcn* is 93.2 GPa and further from *Pbcn* to *Cmcm* 146.1 GPa (Figure S2b). This compares well with experiment [4546] and previous theoretical calculations [42,44,47].

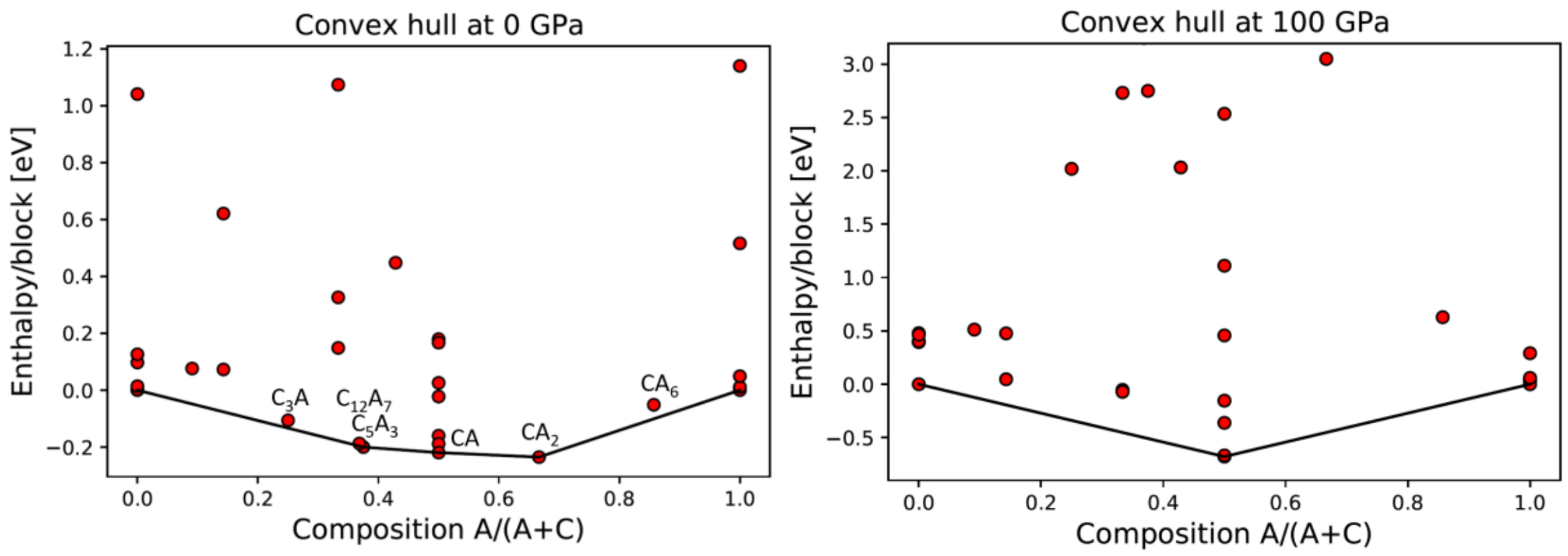

a) b)

Figure 4. Convex hulls of the CaO(C) -$Al_2O_3$ (A) system at 0 GPa (a) and 100 GPa (b).

To find the unknown crystal structure of the "CA-III" phase [13] with cell parameters a=4.390 Å, b=5.070 Å, c=6.960 Å, we performed evolutionary USPEX searches with a fixed $CaAl_2O_4$ and fixed experimental cell parameters. The resulting structure has $P2_12_12$ symmetry, and features layers made of edge- and face-sharing $AlO_6$ octahedra (Figure 5a), with Ca atoms located in the interlayer space. The coincidence of the main X-ray diffraction peaks ((001), (002), (110), (102), (020) and (112)) of our structural model and the experimental [13] diffraction patterns (see Figure 5b) allows us to conclude that the structure was identified correctly. The presence of additional reflections in the theoretical diffraction pattern compared to the experimental one seems likely that the experimental sample was strongly textured, which affected the observed XRD intensities. It should be noted that the predicted crystal structure of the "CA-III" phase is metastable and is 0.09 eV/atom higher in enthalpy than monoclinic $P2_1/n$ phase.

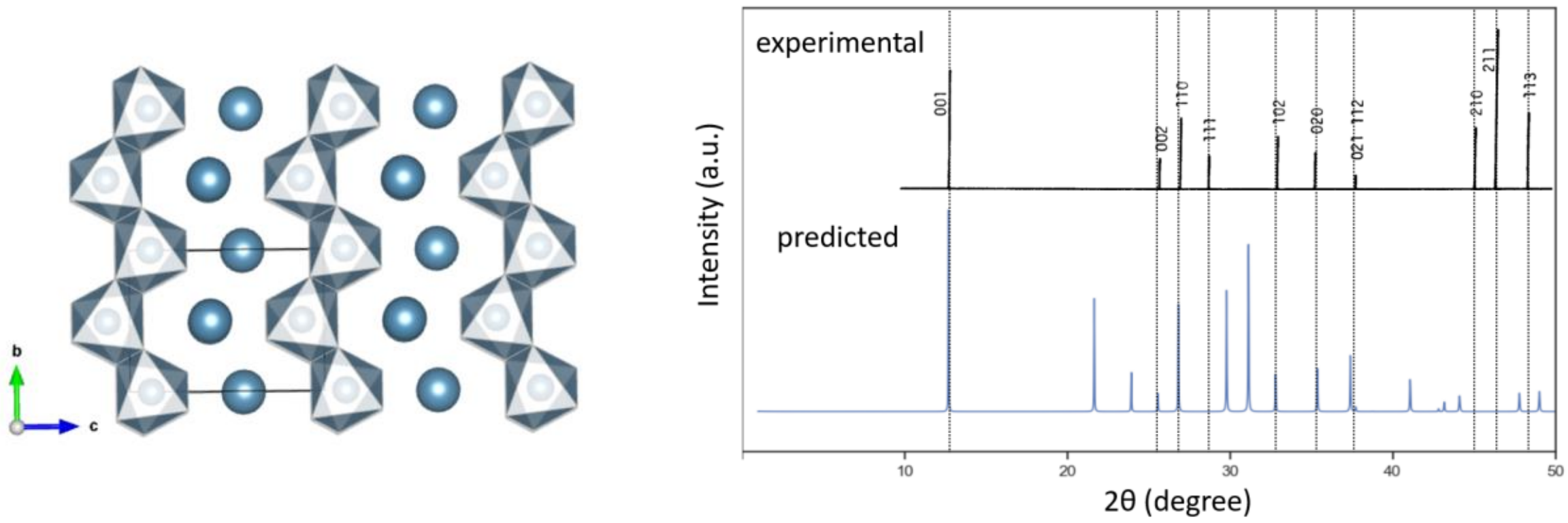

(a) (b)

Figure 5. Predicted structure of "CA-III" metastable phase (a) and diffraction patterns of the "CA-III" phase according to the experimental data from [13] and our predicted structure of "CA-III" phase.

## Conclusions

Our systematic evolutionary searches for stable compounds in the $CaO$-$Al_2O_3$ system at pressures 0-200 GPa clearly establish that at pressures above 7.0 GPa (i.e. at all pressures of the Earth's transition zone and lower mantle) only one calcium aluminate, $CaAl_2O_4$ with $CaFe_2O_4$ structure (*Pnma*), is stable. This phase is one of prime candidate aluminous phases in the lower mantle of the Earth. En route to this conclusion, we have also resolved the old puzzle of the low-pressure (3-8 GPa) metastable phase named "CA-III" – its crystal structure is now established as having the orthorhombic space group $P2_12_12$, and featuring layers of edge- and face-sharing $AlO_6$ octahedra. Our analysis also reveals that at low pressures 5CaO • $3Al_2O_3$ ($C_5A_3$) with space group $Cmc2_1$, $CaAl_4O_7$ ($C2/c$) and $CaAl_2O_4$ ($P2_1/m$) structures are stable up to 2.1, 1.8 and 7.0 GPa respectively.

$CaAl_2O_6$, $CaAl_{12}O_{19}$ and $Ca_{12}Al_{14}O_{33}$ phases are very close to thermodynamic stability at low pressures, and are also known from experiment. It is remain stable that very rich and complex chemistry at low pressures of the cement $CaO$-$Al_2O_3$ system becomes very simple under high pressures.

## Acknowledgments

The research is carried out using Oleg supercomputer of Materials Discovery Lab of Skoltech and the shared research facilities of HPC computing resources at Lomonosov Moscow State University. Work of A.R.O. and E.A.M. is supported by Russian Science Foundation (grant 19-72-30043).

**Stable compounds in the $CaO$-$Al_2O_3$ system at high pressures**

Ekaterina I. Marchenko[1,2], Artem R. Oganov[3], Efim A. Mazhnik[3], Nikolay N. Eremin[1]

[1] – Department of Geology, Lomonosov Moscow State University, GSP-1, 1 Leninskiye Gory, Moscow 119991, Russia.
[2] – Laboratory of New Materials for Solar Energetics, Faculty of Materials Science, Lomonosov Moscow State University, Moscow, Russia
[3] – Skolkovo Institute of Science and Technology, 30 bldg. 1, Bolshoy blvd., Moscow 121205, Russia.

SUPPORTING INFORMATION

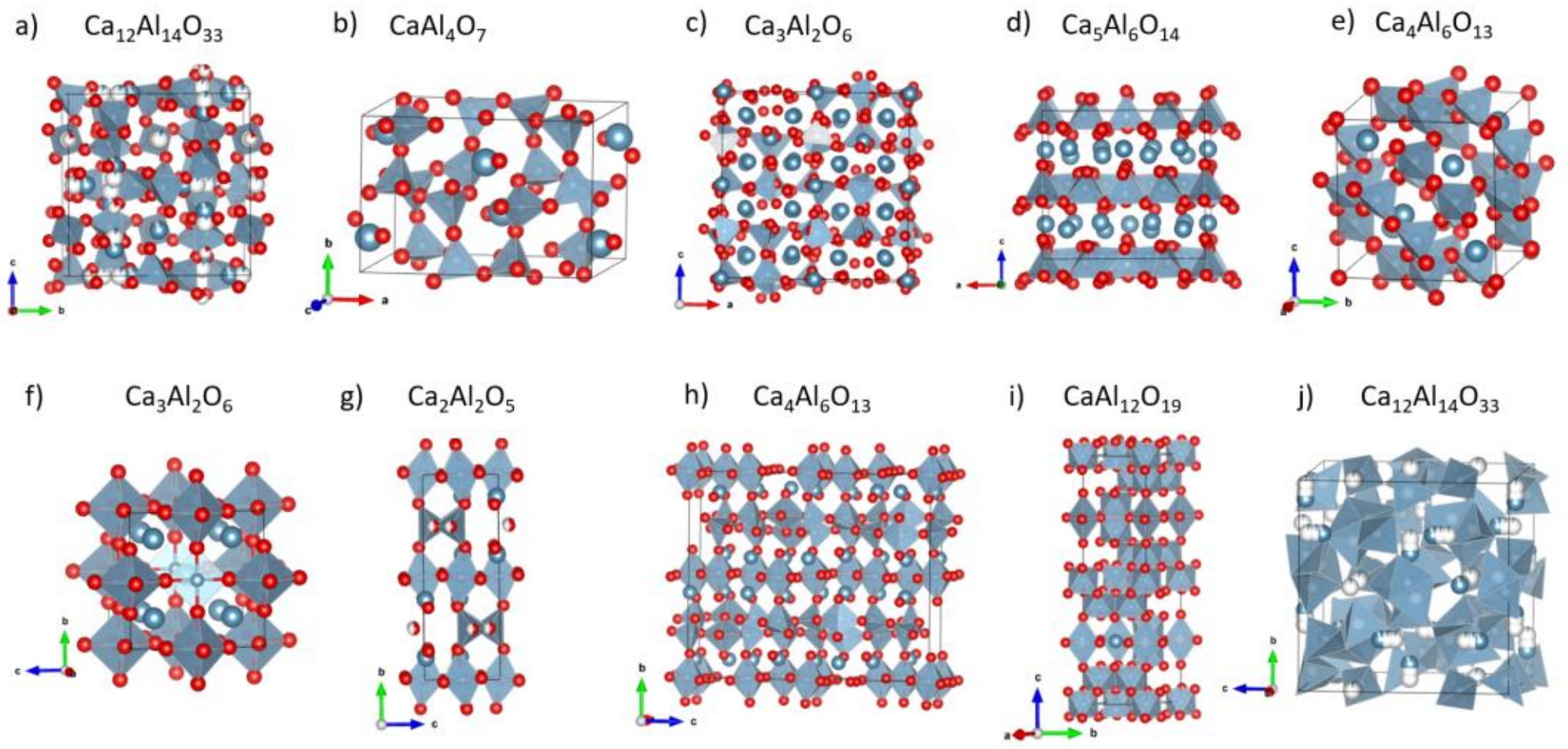

Figure S1. Crystal structures of Ca-Al-O system with different stoichiometric ratios: (a) $Ca_{12}Al_{14}O_{33}$ [20], (b) $CaAl_4O_7$ [21], (c) $Ca_3Al_2O_6$ [22], (d) $Ca_5Al_6O_{14}$ [23], (e) $Ca_4Al_6O_{13}$ [24]), f) $Ca_3Al_2O_6$ [28], (f) $Ca_2Al_2O_5$ [25], (h) $Ca_4Al_6O_{13}$ [27], (i) $CaAl_{12}O_{19}$ [29], and mayenite $Ca_{12}Al_{14}O_{33}$ [30].

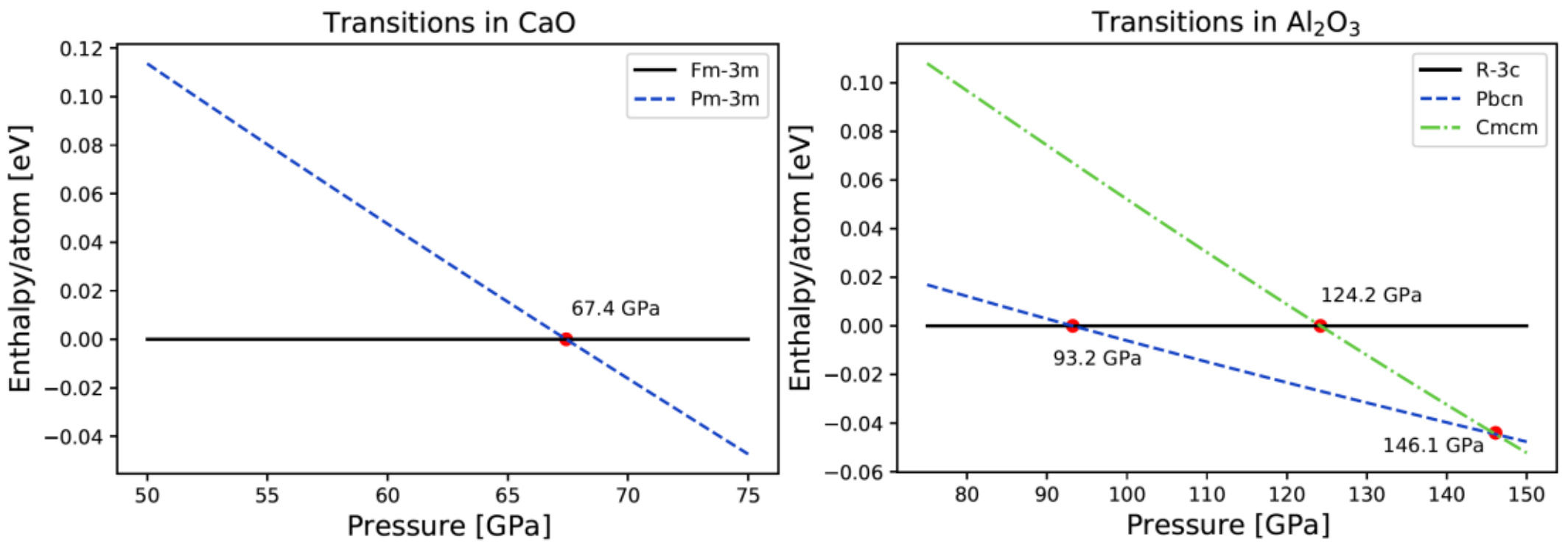

a) b)

Figure S2. The enthalpies of pressure-induced phase transitions in (a) CaO and (b) $Al_2O_3$ relative to $Fm\overline{3}m$ (CaO) and $R\overline{3}c$ ($Al_2O_3$) respectively.

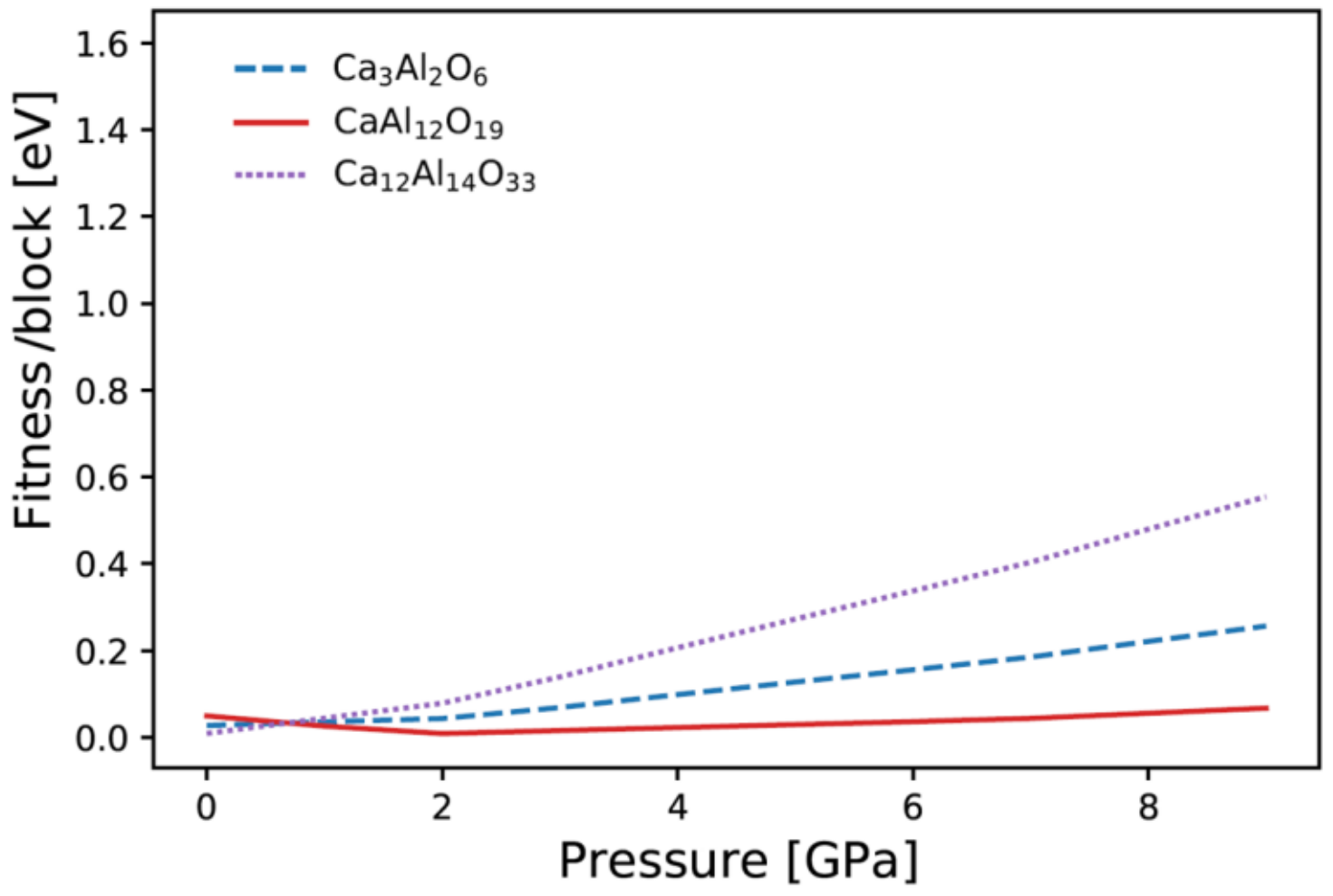

Figure S3. The calculated fitness-pressure graph from convex hull for experimental structures in CaO-$Al_2O_3$ system.